\documentclass{fic-l}

\theoremstyle{definition}

\theoremstyle{remark}

\numberwithin{equation}{section}

\usepackage{graphics}
\begin{document}

\title[Hybrid Stochastic-Deterministic Minimization Method]
{Application of the Hybrid Stochastic-
Deterministic Minimization Method to a
Surface Data
Inverse Scattering Problem
}

\author{S.Gutman}
\address{Department of Mathematics\\ University of Oklahoma\\ Norman,
 OK 73019, USA}
\email{sgutman@ou.edu}
\author{A.G.Ramm}
\address{Department of Mathematics\\
Kansas State University\\ Manhattan, KS 66506-2602, USA}
\email{ramm@math.ksu.edu}


\begin{abstract}
A method for the identification of small inhomogeneities 
from a surface
data is presented in the framework of an inverse scattering
problem for
the Helmholtz equation. 
Using the assumptions of smallness of the scatterers
one reduces this inverse problem to an identification
of the positions of the small scatterers. These positions
are found by a global minimization search. 
Such a search is implemented
by a novel Hybrid Stochastic-Deterministic Minimization method. The
method combines random tries and a deterministic minimization. The
effectiveness of this approach is illustrated by numerical experiments.
In the modeling part our method is valid when the Born
approximation fails. In the numerical part, an algorithm for the
estimate of the number of the small scatterers is proposed. 
 \end{abstract}

\maketitle



\section{Introduction}

In many applications it is essential to find small
inhomogeneities from surface data. For example, such a problem
arises in ultrasound mammography, where small inhomogeneities
are cancer cells. Current X-ray mammography will be replaced by
the ultrasound one because X-ray mammography has a high
probability of creating new cancer cells in a woman's breast in
the course of taking the mammography test.
Other examples include the problem of finding small holes and cracks
in metals and other materials, or the mine detection.
The scattering theory for small scatterers originated in the
 classical works of Lord Rayleigh. It was developed in \cite{R1} and \cite{R2},
 where analytical formulas for the scattering matrix
 were derived for the acoustic and electromagnetic scattering
 problems. In \cite{R1}
and \cite{R3} inverse scattering problems for small bodies are considered.
In \cite{R4} the problem of identification of small subsurface inhomogeneities
from surface data was posed and its possible applications were discussed.
In \cite{KR} the results of the numerical experiments are presented
for the problem of finding one small inhomogeneity from surface data.
In \cite{DR} a method for finding small
inhomogeneities from tomographic data is proposed.
The main result of our paper is the new optimization procedure
which can be used for actual finding the small subsurface inhomogeneities
from surface scattering data. Our numerical results demonstrate that 
the method proposed in this paper is potentially of practical importance.

Consider a point source $y$ of monochromatic acoustic waves on the
surface of the earth. Let $u(x,y,k)$ be acoustic pressure at
the point $x$, and $k>0$ be the wavenumber. The governing equation is:

\begin{equation}
\left[\nabla^2+k^2+k^2v(x)\right] u=-\delta(x-y)\ \text{in}\ R^3,
\end{equation}
  
$u$ satisfies the radiation condition at infinity, and $v(x)$
is the inhomogeneity in the velocity profile, $x=(x_1,x_2,x_3)$.

Let us assume that $v(x)$ is a bounded function vanishing outside of the domain
$D=\cup_{m=1}^M D_m $
which is the union of $M$ small nonintersecting domains $D_m$,
all of them are located in the lower half-space
$R^3_{-}=\{x:x_3<0\}$. Smallness is understood in the sense
$k\rho\ll 1$, where $\rho:=\frac 12 \max_{1\leq m\leq
M}\{\text{diam}\,D_m\}$,
and diam $D$ is the diameter of the domain $D$. Practically
$k\rho\ll 1$ means that $k\rho<0.1$, in some cases $k\rho<0.2$ is
sufficient
for obtaining acceptable numerical results. The background velocity
in (1.1) equals to 1, but we can consider the case of fairly
general background velocity \cite{R3}.

Denote $\tilde z_m$ and $\tilde v_m$ the positions of the gravity
center of $D_m$ and the total intensity of the $m\text{-th}$
inhomogeneity $\tilde v_m:=\int_{D_m}v(x)dx$. Assume that
$\tilde v_m\not= 0$.

The inverse problem to be solved is:

{\bf IP:} {\it Given $u(x,y,k)$ for all $(x,y)$ on
$P:=\{x:x_3=0\}$ at a fixed $k>0$, find
the number $M$ of small inhomogeneities,
the positions $\tilde z_m$ of the inhomogeneities,
and their intensities $\tilde v_m$.
}

\section{Method of solution}

Let us introduce the following notations:
\begin{equation}
P:=\{x:x_3=0\},
\end{equation}

\begin{equation}
\{x_j,y_j\}:=\xi_j,\quad 1\leq j\leq J,\quad x_j,y_j\in P, 
\end{equation}

are the points at which the data
   $\  u(x_j,y_j,k)\ $are collected,
\begin{equation}
k>0\ \text{is fixed}, 
\end{equation}

\begin{equation}
g(x,y,k):=\frac{\exp(ik|x-y|)}{4\pi|x-y|},
\end{equation}

\begin{equation}
G_j(z):=G(\xi_j,z):=g(x_j,z,k)g(y_j,z,k),
\end{equation}

\begin{equation}
f_j:=\frac{u(x_j,y_j,k)-g(x_j,y_j,k)}{k^2},
\end{equation}

\begin{equation}
 \Phi(z_1,\dots,z_M,\, v_1,\dots,v_M):=
   \sum^J_{j=1}\left| f_j-\sum^M_{m=1} G_j(z_m)v_m\right|^2.
\end{equation}

The proposed method for solving the (IP) consists in finding the
global minimizer of function (2.7). This minimizer
$(\tilde z_1,\dots,\tilde z_M,\, \tilde v_1,\dots,\tilde v_M)$
gives the estimates  of the
positions $\tilde z_m$ of the small inhomogeneities and their intensities
$v_m$.

The above approach can be justified as follows.
The Helmholtz equation (1.1) with the radiation condition
is equivalent to the integral equation
\begin{equation}
u(x,y,k)=g(x,y,k)+k^2 \sum^M_{m=1}\int_{D_m}
   g(x,z,k)v(z)u(z,y,k)dz.
\end{equation}

For small inhomogeneities the integral on the right-hand side of
(2.8) can be approximately written as
\begin{multline}
 \int_{D_m} g(x,z,k)v(z)u(z,y,k)dz
    \approx \int_{D_m}g(x,z,k)v(z)g(z,y,k)dz \\
  \approx G(x,y,\overline z_m)
   \int_{D_m}vdz:=G(\xi,\overline z_m)\tilde v_m,\quad 1\leq m\leq M,  
\end{multline}

where $\overline z_m$ is a point close to $\tilde z_m$, and
$u(z,y,k)$ under the sign of the integral in (2.8) can be replaced
by $g(x,y,k)$ with a small error $\varepsilon^2$ provided that
\begin{equation}
\varepsilon:=c_0M(k\rho)^2\ll 1, 
\end{equation}

where
\begin{equation}
 \rho=\max_{1\leq m\leq M} \rho_m, \quad \rho_m:=\frac 12 diam D_m,\quad
   c_0:=\max_{x\in{\Bbb R}^3} |v(x)|, 
\end{equation}

and $M$ is the number of inhomogeneities.

Note that the sufficient condition for the
validity of the  Born approximation for equation
(2.8),  is the smallness of the norm in $L^2(D)$
of the integral operator in (2.8).
The above condition can be written as  
$$ M c_0 k^2 \rho^2 <<1. \quad (*)
$$ 
If $x,y\in P$, then 
$$|u-g|<O(Mk^2c_0\rho^3 d^{-2}),$$
where $d>0$ is the minimal distance from $D_m$ to $P$. 

The relative error of the first approximation in
(2.9) for $x,y\in P$ is $O(M k^2c_0\rho^3 d^{-1}),$
where $d>0$ is the minimal distance from $D_m$ to $P$. 
If $d>0$ is not too small, 
$c_0\rho^3:=V$ is not too large, then the above error
is small if  
$$\epsilon_1:=M k^2c_0\rho^3 d^{-1}<<1, \quad (**) 
$$ 
where
$\epsilon_1$ is a dimensionless quantity.
Condition (*) can be written as $\frac {M k^2 V} {\rho} <<1$.
This condition is much stronger than condition (**), which
can be written as $\frac {k^2 V}{d}<<1$, since $d>>\rho$.
Therefore the approximation (2.9) is applicable when the Born
approximation may fail.

If the condition  
$$c_0\rho^3=const\quad \text {as}\quad \rho \to 0 \quad (***)$$
holds, then condition (*) for the validity of the Born approximation
is violated because $\frac {M k^2 V} {\rho}\to \infty$ as 
$\rho \to 0.$

Condition (***) corresponds to the scattering by delta-type
inhomogeneities. For such scattering the Born approximation is not
applicable, but a modified representation of the type (2.9) is valid
(see \cite {AG}, p.113, formula (1.1.33), and \cite{GR}, \cite{R5}).

From (2.9) and (2.6) it follows that
\begin{equation}
 f_j\approx\sum^M_{m=1}G_j(\overline z_m)v_m, \quad
   G_j(\overline z_m):=G(\xi_j,\overline z_m,k) 
\end{equation}

Therefore, parameters $\tilde z_m$ and $\tilde v_m$ can be
estimated by the least-squares method if one finds the global
minimum of the function (2.7):
\begin{equation}
\Phi(z_1,\dots,z_M,\, v_1,\dots,v_M)=\min.
\end{equation}

The function $\Phi$ depends on $M$ unknown points
$z_m\in{\Bbb R}^3_-$, and $M$ unknown parameters $v_m$,
$1\leq m\leq M$. The parameter $M$, the number of the small
inhomogeneities, is estimated by the algorithm described in Sec. 3 below,
see also a discussion in the beginning of Section 4.

\section{Hybrid Stochastic-Deterministic Method}

Let the inhomogeneities be located within a box

\begin{equation}
B=\{(x_1,x_2,x_3)\ :-a<x_1<a,\ -b<x_2<b,\ 0<x_3<c\}\,,
\end{equation}
and their intensities satisfy 

\begin{equation}
0\leq v_m\leq v_{max}\,.
\end{equation}

Then, given the location of the points $z_1, z_2, \dots, z_M$, the
minimum of $\Phi$ in (2.13) with respect to the intensities $v_1, v_2,
\dots, v_M$ can be found by minimizing the quadratic function in (2.7) over the region
satisfying (3.2). This can be done using normal equations for (2.7) and
projecting the resulting point back onto the region defined by (3.2).
Denote the result of this minimization by $\tilde \Phi$, that is

\begin{equation}
\begin{split}
\tilde \Phi(z_1, z_2, \dots, z_M)& =\min\{\Phi(z_1, z_2, \dots, z_M,
v_1, v_2, \dots, v_M)\ :\\
&\ 0\leq v_m\leq v_{max}\,,\quad 1\leq m \leq M\}
\end{split}
\end{equation}

Now the minimization problem (2.13) is reduced to the $3M$-dimensional
restrained minimization
\begin{equation}
\tilde \Phi(z_1, z_2, \dots, z_M)=\min \,,\quad z_m\in B\,,\quad 1\leq m
\leq M\,.
\end{equation}

Note, that the dependency of $\tilde\Phi$ on its $3M$ variables (the
coordinates of the points $z_m$) is highly nonlinear. In particular,
this dependency is complicated by the computation of the minimum in
(3.3) and the consequent projection onto the admissible set $B$. Thus,
an analytical computation of the gradient of $\tilde\Phi$ is not
computationally efficient. We have used Powell's quadratic minimization
method to find local minima. This method uses a special procedure to numerically approximate the
gradient, and it can be shown to exhibit the same type of quadratic
convergence as conjugate gradient type methods (see \cite{Bre}).

In addition, 
{\it the exact number of the original inhomogeneities $M_{orig}$ is
unknown,
and its estimate is a part of the inverse problem.} In the HSD algorithm
described below this task is accomplished by taking an initial number
$M$ sufficiently large, so that

\begin{equation}
M_{orig}\leq M\,,
\end{equation}
which, presumably, can be estimated from physical considerations. After
all, our goal,  is to find only the strongest inclusions, since
the weak ones cannot be distinguished from the background noise. The
Reduction Procedure (see below) allows the algorithm to seek the minimum
of $\tilde\Phi$ in a lower dimensional subsets of the admissible set
$B$, thus finding the estimated number of inclusions $M$. Still another
difficulty in the minimization is a large number of local minima of
$\tilde \Phi$. While this phenomena is well known for the objective
functions arising in various inverse problems, we illustrate this point
on Figure 1. 

\begin{figure}[tb]
\vspace{5pc}
\includegraphics*{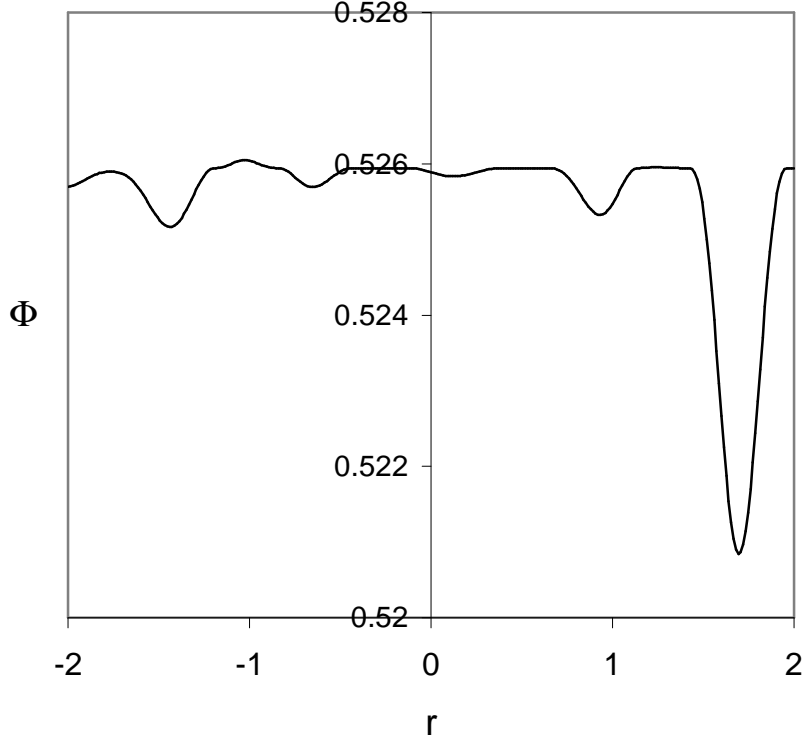}
\caption{Objective
function $\tilde\Phi(z_r,z_2,\tilde z_3,\tilde z_4,\tilde z_5,\tilde
z_6)\,,\quad -2\leq r\leq 2$ }
\end{figure}

The original configuration 
$(\tilde z_1, \tilde z_2, \dots, \tilde z_M,
\tilde v_1, \tilde v_2, \dots, \tilde v_M)\,,\; M_{orig}=6$
is presented in Table 1. Figure 1 shows the values of the 
function $\tilde\Phi(z_r,z_2,\tilde z_3,\tilde z_4,\tilde z_5,\tilde
z_6)$, where
\[
z_r=(r,0,0.520)\,,\quad -2\leq r\leq 2
\]
and
\[z_2=(-1,0.3,0.580)\,.
\]
The plot shows multiple local minima and almost flat regions. 

A direct application of a gradient type method to such a function would
result in finding a local minimum, which may or may not be the sought
golobal one. In the example above, such a method would usually be
trapped in a local minimum located at $r=-2,\ r=-1.4,\ r=-0.6,\ r=0.2$
or $r=0.9$. While the desired global minimum at $r=1.6$ will be found
only for a sufficiently close initial guess $1.4<r<1.9$. While various
global minimization methods are known (see below), we found that the most
efficient way to accomplish the task for this Inverse Problem was to
design a new method (HSD) combining both the
stochastic and the deterministic approach to the global minimization.
Deterministic minimization algorithms with or without the gradient
computation, such as the conjugate gradient methods, are known to be
highly efficient (see \cite{Bre,DeS,Pol,Jac}). However, the 
initial guess should be chosen
sufficiently close to
the sought minimum.  Also such algorithms, as we mentioned above,  tend to be trapped
at a local minimum, which is not necessarily close to a global one.
A new deterministic method is proposed in 
\cite{Bar} and \cite{BPR}, which is quite
efficient according to \cite{BPR}.
On the other hand, various stochastic minimization algorithms, e.g. the
simulated annealing method \cite{KGV,Kir}, are more likely to find a 
global minimum,
but their convergence can be very slow. We have tried a variety of minimization
algorithms to find an acceptable minimum of (3.4) or (2.13). Among them were the 
Levenberg-Marquardt Method, Conjugate Gradients,
Downhill Simplex, and Simulated Annealing Method. None of them
produced consistent satisfactory results. 

Among the minimization methods combining random and deterministic searches
we mention Deep's method \cite{Deep} and a variety of clustering methods
\cite{RKT1}, \cite{RKT2}. An application of these methods to
the particle identification using light scattering is 
described in \cite{ZUB}. The clustering methods are quite robust but, usually,
require a significant computational effort. The HSD method is a combination of 
a reduced sample random search method with certain 
ideas from Genetic Algorithms (see e.g. \cite{HH}).
It is very efficient and seems especially well suited for low dimensional global
minimization. Further research is envisioned to study its 
properties in more detail, and its applicability to other problems.

While the steps of the Hybrid Stochastic-Deterministic
 (HSD) method are outlined below,
its basic idea is to
start with a random search for the minimum.
 If at a certain configuration $x_s$ the
minimized function is judged to be sufficiently small, this
configuration is a candidate for the initial 
guess of a deterministic minimization method.
However, this minimization is applied only after 
the points with smaller intensities in $x_s$
are dropped (Step 2 below), and all sufficiently
 close points are eliminated (Step 3).
Now a deterministic minimization is performed 
in the resulting lower dimensional subspace.
The minimizer 
 $x_d$ is likely to represent at least some of
the sought inhomogeneities. 
These points are supplemented with
 randomly chosen ones to obtain a full configuration
(Step 1).
In this manner
the iterations of the random searches
and the deterministic minimization continue till a tolerance criterion
is satisfied. As the algorithm finds configurations with smaller and
smaller values of the objective function, the likelihood of the finding
 the global minimum increases, while the irrelevant local minima are
eliminated.

Let us outline the steps of a possible 
implementation of the HSD algorithm for the surface data inverse
scattering problem described in Sections 1 and 2.
We assume that all the inhomogeneities are located in a (known) box $B$,
and
$M$ is chosen according to (3.5).
Our goal is to locate the points $\tilde z_1,\dots,\tilde z_N,\; N\leq M$ which
minimize $\tilde\Phi$ in (3.4).

\subsection*{Hybrid Stochastic-Deterministic (HSD) Method}

Let $P_0, T_{max}, \epsilon_s, \epsilon_i, \epsilon_d$, and $\epsilon$ be positive numbers. 
 Let $N=0$.

\begin{enumerate}

\item   Generate
randomly
$M-N$ additional points $z_{N+1},\dots,z_M$ to obtain a full
configuration $z_1,\dots,z_M$. Compute $P_s=\tilde\Phi(z_1,z_2 \dots, z_M)$.
Save the resulting best fit intensities $v_m\,,\; 1\leq m\leq M$ (see 3.3).
If $P_s<P_0\epsilon_s$ then go to step 2, otherwise repeat this
step 1.

\item  Discharge all the points with the intensities $v_m$ satisfying $v_m<v_{max}\epsilon_i$. 
Now only $N\leq M$ points $z_1,z_2 \dots, z_N$ remain in the box $B$.

\item If any two points $z_m, z_n$ in the above configuration satisfy 
$|z_m-z_n|<\epsilon_dD$, where $D=diam(B)$, then eliminate point $z_n$,
change the intensity of point $z_m$ to $v_m+v_n$, and assign $N:=N-1$.
This step is repeated until no further reduction in $N$ is possible.

\item Run a restrained deterministic minimization of $\tilde\Phi$
in $3N$ variables, with the
initial guess at the configuration determined in step 3. Let the minimizer be
$\tilde z_1,\dots,\tilde z_N$.
If $P=\tilde\Phi(\tilde z_1,\dots,\tilde z_N)<\epsilon$ then save this
configuration,
and go to step 6, otherwise let $P_0=P$, and
proceed to the next step 5.

\item Keep intact N points $\tilde z_1,\dots,\tilde z_N$. 
If the number of random
configurations has exceeded $T_{max}$ (the maximum
number of random tries), then save the configuration
 and go to step 6, else go to step 1, and use these $N$ points there.

\item  Repeat steps 1 through 5 $n_{max}$ times. 

\item Find the configuration among the above $n_{max}$ ones, which gives the smallest
value to $\tilde\Phi$. This is the best fit.

\end{enumerate}

Step 1 is the stochastic part of the algorithm. In step 2 the
points with low intensities are discharged, since they, most likely, are
 artifacts contributing on the level comparable with the background noise.
Step 3 is the Reduction Procedure replacing two nearby inclusions with
one of the joint intensity. Steps 2 and 3 lower the dimensionality of the
minimization domain, thus greatly reducing the computational time needed
to perform the deterministic minimization of the step 4. 
We have used Powell's minimization method
(see \cite{Bre} for a detailed description) for the deterministic part, since
this method does not need
gradient computations and converges quadratically near quadratically
shaped minima. Also, in step 1, an idea from the Genetic Algorithm's approach \cite{HH} is
implemented by keeping only the strongest representatives of the
population and allowing a mutation for the rest.

\section{Numerical results}

The algorithm was tested on a variety
of configurations. Here
we present the results of just
two typical numerical experiments illustrating the performance of the
method. The data was simulated according to (2.12).
In both experiments the box $B$ is taken to be
$$B=\{(x_1,x_2,x_3)\ :-a<x_1<a,\ -b<x_2<b,\ 0<x_3<c\}\,,$$
with $a=2,\ b=1,\ c=1$. The frequency $k=5$, and the effective
intensities $v_m$ are in the range from $0$ to $2$, (see Section 1).
The values of the parameters defined in Section 3 were chosen as follows
$$P_0=1\,,T_{max}=1000,\,\epsilon_s=0.5,\,\epsilon_i=0.25,\,\epsilon_d=0.1\,,
\epsilon=10^{-5}\,,
n_{max}=6$$
In both cases we searched for the same 6 inhomogeneities with the
 coordinates $x_1,x_2,x_3$ and the intensities $v$ shown in Table 1.

\begin{table}
\caption{Actual inclusions.}
\begin{tabular}{|r|r|r|r|r|}
\hline
Inclusions &    $\qquad   x_1$ & $\qquad x_2$ & $\qquad x_3$ & $\qquad v$ \\
\hline
1 & 1.640 & -0.510 & 0.520 & 1.200 \\
2 & -1.430 & -0.500 & 0.580 & 0.500 \\
3 & 1.220 & 0.570 & 0.370 & 0.700 \\
4 & 1.410 & 0.230 & 0.740 & 0.610 \\
5 & -0.220 & 0.470 & 0.270 & 0.700 \\
6 & -1.410 & 0.230 & 0.174 & 0.600 \\
\hline
\end{tabular}

\end{table} 

Parameter $M$ was set to 16, thus the only information on the number of
inhomogeneities given to the algorithm was that their number does not
exceed 16. This number was chosen to keep the computational time within
reasonable limits. Still another consideration for the number $M$ is the
aim of the algorithm to find the presence of the most influential
inclusions, rather then all inclusions, which is usually impossible in
the presence of noise and with the limited amount of data.

{\bf Experiment 1.}
In this case we used 12 sources and 21 detectors, all on the surface
$x_3=0$.
The sources were positioned at $\{(-2+0.333+0.667i,-0.5+1.0j,0),\
i=0,1,\dots,5,\,j=0,1\}$, that is 6 each along two lines $x_2=-0.5$ and
$x_2=0.5$.
The detectors were positioned at $\{(-2+0.667i,-1.0+1.0j,0),\
i=0,1,\dots,6,\,j=0,1,2\}$, that is seven detectors along each of the three lines
$x_2=-1,\, x_2=0$ and
$x_2=1$.
This corresponds to a mammography
search, where the detectors and the sources are placed above the search
area. The results of the identification are shown in Tables 2 and 3
for different noise levels $\delta$ in the data. See Figure 2. 
Table 3 has only 4 lines showing that the program has identified
the presence of only 4 indicated inclusions, while missimg the other 2.

To
evaluate the performance of the algorithm, we have made 10
independent runs of the program for each of the noise levels
$\delta=0.00,\ \delta=0.02$, and $\delta=0.05$. In the noiseless case
all 6 original inhomogeneities were identified every time. The program
performed 2 or 3 deterministic minimizations in each run, spending from
30 to 80 percent of time in this minimization. For $\delta=0.02$ the
perfect identification also was obtained in every run. The deterministic
minimization was performed 2-4 times, spending there 10 to 25\% of the
computational time. For $\delta=0.05$ out of 10 runs the program found 4
and 5 inhomogeneities in one run each, and all 6 inhomogeneities in the rest 8 runs.
The deterministic minimization was performed from 1 to 3 times in each run for the
total of 5 to 20\% of the computational time.

\begin{table}
\caption{Experiment 1. Identified inclusions, no noise, $\delta=0.00$.}
\begin{tabular}{|r|r|r|r|}
\hline
 $\qquad   x_1$ & $\qquad x_2$ & $\qquad x_3$ & $\qquad v$ \\
\hline
 1.640 & -0.510 & 0.520 & 1.20000 \\
 -1.430 & -0.500 & 0.580 & 0.50000 \\
  1.220 & 0.570 & 0.370 & 0.70000 \\
 1.410 & 0.230 & 0.740 & 0.61000 \\
 -0.220 & 0.470 & 0.270 & 0.70000 \\
-1.410 & 0.230 & 0.174 & 0.60000 \\
\hline
\end{tabular}

\end{table}

\begin{table}
\caption{Experiment 1. Identified inclusions, $\delta=0.05$.}
\begin{tabular}{|r|r|r|r|}
\hline
 $\qquad   x_1$ & $\qquad x_2$ & $\qquad x_3$ & $\qquad v$ \\
\hline
 1.645 & -0.507 & 0.525 & 1.24243 \\
 1.215 & 0.609 & 0.376 & 0.67626 \\
  -0.216 & 0.465 & 0.275 & 0.69180 \\
-1.395 & 0.248 & 0.177 & 0.60747 \\
\hline
\end{tabular}

\end{table}

\begin{figure}[tb]
\vspace{5pc}
\includegraphics*{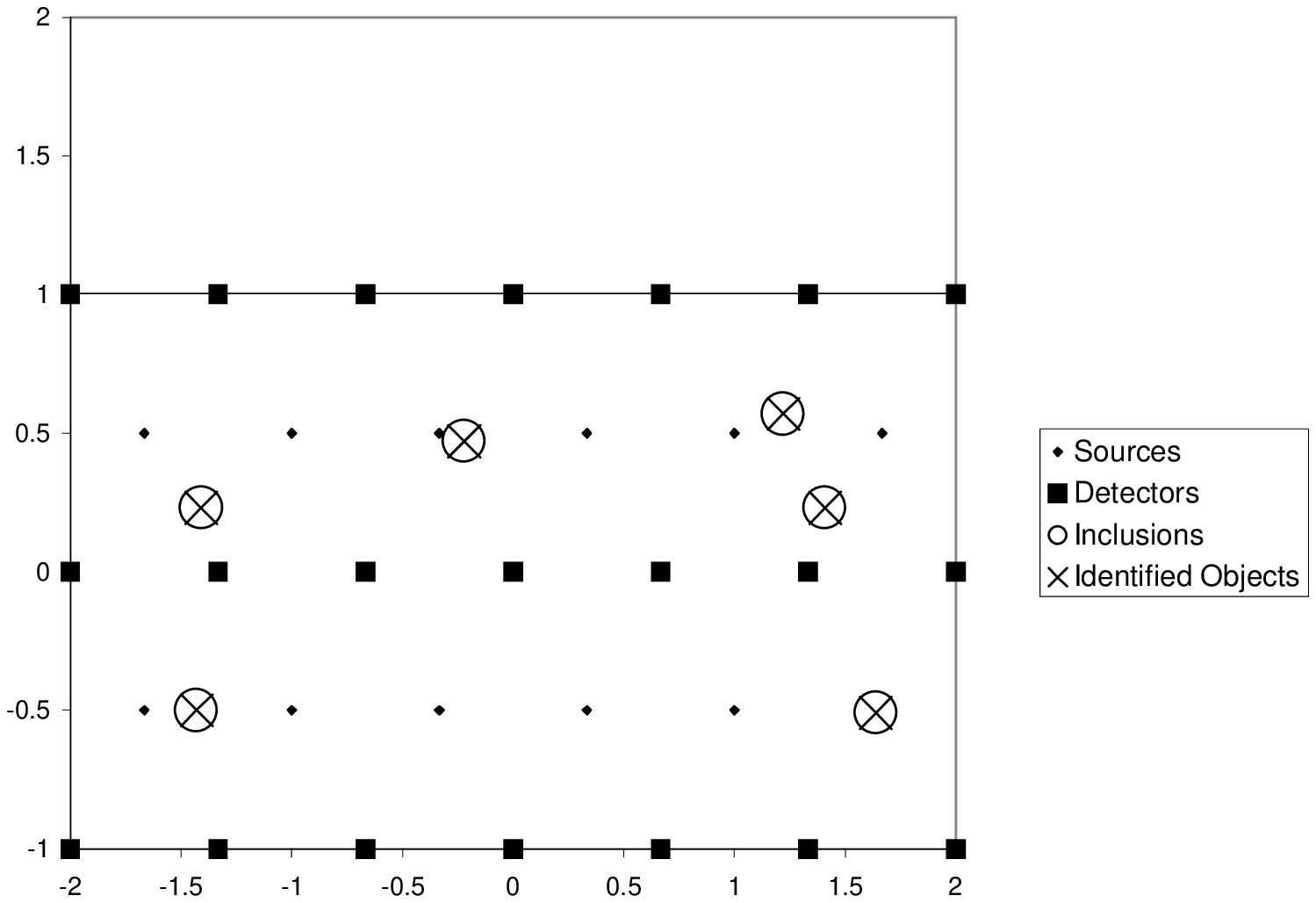}
\caption{Inclusions and Identified objects for Experiment 1, $\delta=0.00$. $x_3$ coordinate is not shown.}
\end{figure}

{\bf Experiment 2.}
In this case we used 8 sources and 22 detectors, all on the surface
$x_3=0$.
The sources were positioned at $\{(-1.75+0.5i,1.5,0),\
i=0,1,\dots,7,\,j=0,1\}$, that is all 8 along the line $x_2=1.5$.
The detectors were positioned at $\{(-2+0.4i,1.0+1.0j,0),\
i=0,1,\dots,10,\,j=0,1\}$, that is eleven detectors along each of the two lines
$x_2=1$ and
$x_2=2$.
This corresponds to a mine
search, where the detectors and the sources must be placed outside of
the searched ground. The results of the identification are shown in Tables 4 and 5
for different noise levels $\delta$ in the data. See Figure 3. Because
the sources and the detectors are further away from the inhomogeneities
than in Experiment 1, this presents a more difficult identification
problem. The program identified less than 6 original inhomogeneities in
some runs, thus Tables 4 and 5 contain less than 6 lines. As it can be
seen, only the strongest inhomogeneities have been identified in these
cases. 

As in Experiment 1, 
we have made 10
independent runs of the program for each of the noise levels
$\delta=0.00,\ \delta=0.02$, and $\delta=0.05$. In the noiseless case
$\delta=0.00$,
5 or 6 original inhomogeneities were identified every time. In 3 runs an additional
inhomogeneity (an artifact) has appeared. 
For $\delta=0.02$ and  $\delta=0.05$ the identification has
deteriorated, with 4 inhomogeneities recovered. The number of
deterministic minimizations and the computational times spent doing them
are somewhat higher than the ones reported for Experiment 1.

\begin{table}
\caption{Experiment 2. Identified inclusions, no noise, $\delta=0.00$.}
\begin{tabular}{|r|r|r|r|}
\hline
 $\qquad   x_1$ & $\qquad x_2$ & $\qquad x_3$ & $\qquad v$ \\
\hline
 1.656 & -0.409 & 0.857 & 1.75451 \\
 -1.476 & -0.475 & 0.620 & 0.48823 \\
 1.209 & 0.605 & 0.382 & 0.60886 \\
  -0.225 & 0.469 & 0.266 & 0.69805 \\
-1.406 & 0.228 & 0.159 & 0.59372 \\
\hline
\end{tabular}

\end{table} 

\begin{table}
\caption{Experiment 2. Identified inclusions, $\delta=0.05$.}
\begin{tabular}{|r|r|r|r|}
\hline
 $\qquad   x_1$ & $\qquad x_2$ & $\qquad x_3$ & $\qquad v$ \\
\hline
 1.575 & -0.523 & 0.735 & 1.40827 \\
 -1.628 & -0.447 & 0.229 & 1.46256 \\
 1.197 & 0.785 & 0.578 & 0.53266 \\
 -0.221 & 0.460 & 0.231 & 0.67803 \\
\hline
\end{tabular}

\end{table}

\begin{figure}[tb]
\vspace{5pc}
\includegraphics*{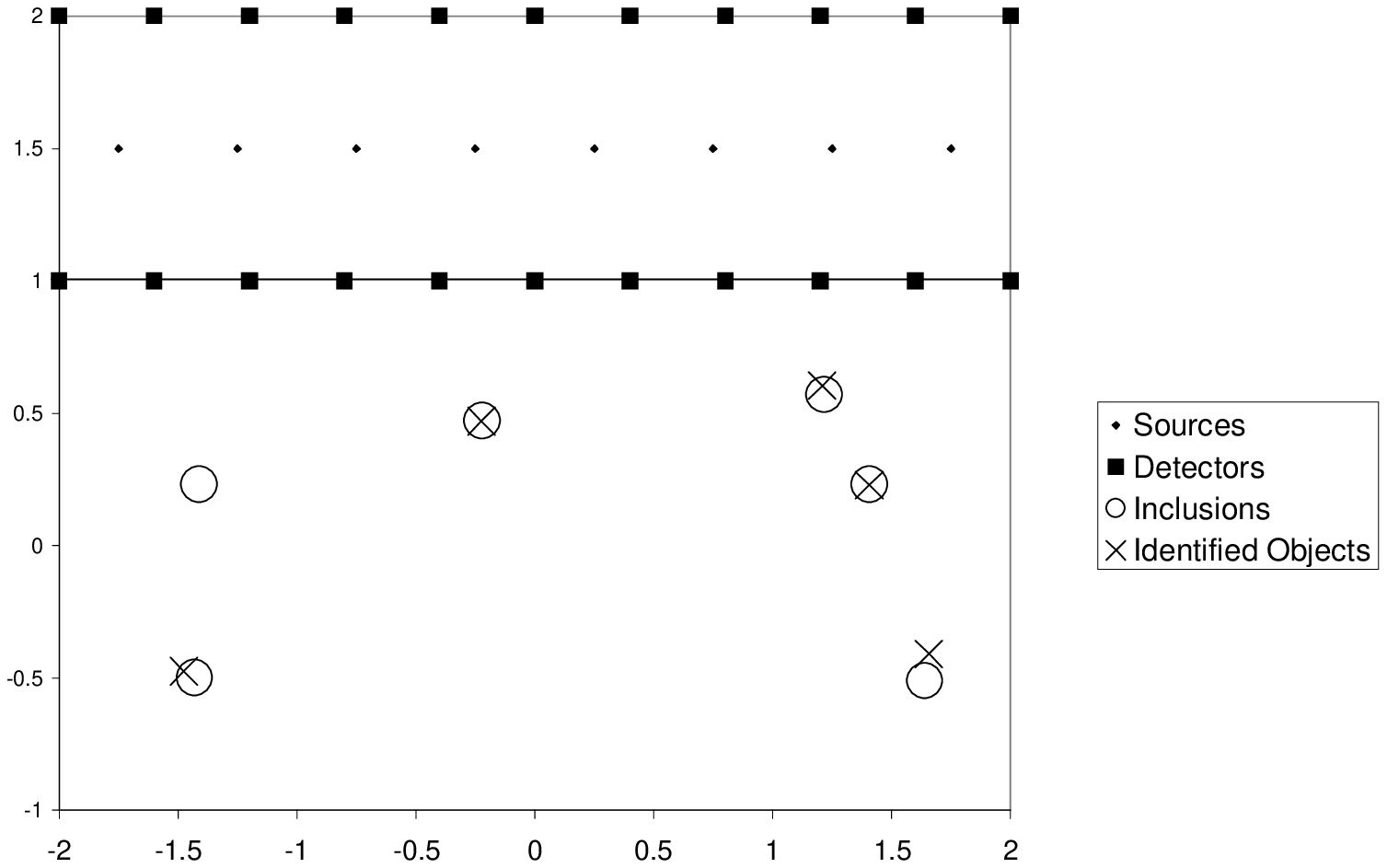}
\caption{Inclusions and Identified objects for Experiment 2, $\delta=0.00$. $x_3$ coordinate is not shown.}
\end{figure}

In general, the execution times were less than 2 minutes on a 333MHz PC. As it can be
seen from the results, the method achieves a perfect identification in
the Experiment \#1 when no noise is present. The identification
deteriorates in the presence of noise, as well as if the sources and
detectors are not located directly above the search area. Still the
inclusions with the highest intensity and the closest ones to the surface
are identified, while the deepest and the weakest are lost. This can be
expected, since their influence on the cost functional is becoming
comparable with the background noise in the data.

In summary, the proposed method for the identification of small inclusions
can be used in geophysics, medicine and technology.
It can be useful in the development of new approaches to ultrasound
mammography. It 
can also be used  
for localization of holes and cracks in metals and other
materials, as well as for finding mines from
surface measurements of acoustic pressure and possibly in other
problems of interest in various applications.

The HSD minimization method is a specially designed low-dimensional
minimization method, which is well suited for many inverse type
problems. The problems do not necessarily have to be within the Born approximation
range. It is highly desirable to study applications of this method, and
to compare its performance to other competitive methods.


S.Gutman,
Department of Mathematics,
University of Oklahoma,
Norman, OK 73019, USA,
e-mail: sgutman@ou.edu

A.G.Ramm,
Department of Mathematics,
Kansas State University,
Manhattan, KS 66506-2602, USA,
e-mail: ramm@math.ksu.edu

\end{document}